\documentclass[UTF8,a4paper]{article}
\usepackage{amsmath}
\usepackage{graphicx}
\usepackage{subfigure}
\usepackage{epstopdf}
\usepackage{geometry}
\usepackage{ulem}
\usepackage{indentfirst}
\usepackage{amsfonts,amssymb,dsfont}
\usepackage{setspace}
\usepackage{threeparttable}
\usepackage{amsmath,mathtools,amsthm}
\usepackage{array}
\usepackage{extarrows}
\usepackage{upgreek}

\makeatletter
\renewcommand*\env@matrix[1][\arraystretch]{%
  \edef\arraystretch{#1}%
  \hskip -\arraycolsep
  \let\@ifnextchar\new@ifnextchar
  \array{*\c@MaxMatrixCols c}}
\makeatother
\begin{document}


\title{\bf Electronic transport and dynamical polarization in bilayer silicene-like system}
\author{Chen-Huan Wu
\thanks{chenhuanwu1@gmail.com}
\\College of Physics and Electronic Engineering, Northwest Normal University, Lanzhou 730070, China}

\maketitle
\vspace{-30pt}
\begin{abstract}
\begin{large} 

We investigate the semiclassical electronic transport properties of the bilayer silicene-like system
in the presence of charged impurity.
The trigonal warping due to the interlayer hopping, and its effect to the band structure of bilayer silicene is discussed.
Besides the trigonal warping, the external field also gives rise to the anisotropic effect 
of the mobility (at finite temperature)
which can be explored by the Boltzmann theory.
The dynamical polarization as well as the scattering wave vector-dependent screening
within random phase approximation are very important in determining the 
scattering behavior and the self-consistent transport.
We detailly discuss the transport behavior under the 
short- or long-range potential.
The phonon scattering with the acoustic phonon mode which dominant at high temperature is also studied
within the density functional theory (DFT).
Our results are also valid for the bilayer graphene or bilayer MoS$_{2}$,
and other bilayer systems with strong interlayer coupling.\\




\end{large}

\end{abstract}
\begin{large}

\section{Introduction}

The bilayer silicene-like group-IV two-dimension (2D) materials, like the bilayer graphene has been widely studied and compared 
with the silicene\cite{Ohta T}.
Except the group-IV materials,
the electronic transport properties of the bilayer MoS$_{2}$ (with three sublayers) as well as the bilayer blue phosphorene\cite{Ospina D A}
(which has a structure more similar to the silicene\cite{Jain A} than the bilayer black phosphorene) 
also has been compared with the silicene, including some of the thermodynamics properties, like the
thermal conductivity\cite{Jain A} and the temperature-dependent scattering (and with the similar phonon spectrum).
The MoS$_{2}$ and the blue phosphorene also have a buckled structure\cite{Wu C H1,Kim S J,Liu Y} 
due to the $sp^{2}-sp^{3}$ hybridization 
like the silicene.
The optical properties of the bilayer silicene-like parabolic system, 
including the absorption of the radiation (by the polarized light\cite{John R}), 
also has been widely studied\cite{Yang L}.


The properties including the dynamical polarization between the linear Dirac system (monolayer silicene) and the parabolic system
(bilayer silicene)
has a large difference.
For band structure of moolayer silicene,
the symmetry between the conduction band and the valence band
will be broken in silicene by the next-nearest-neighbor (NNN) Rashba-coupling
(induced by the applied perpendicular electric field) (see Refs.\cite{Wu C H3,Wu C HX,Scholz A2}).
We found that, the static polarization of bilayer silicene could be nearly temperature-independent in adiabatically case
with the large band gap.
In addition, the difference in band gap also affects the mobility and the phonon scattering as well as
the Landau level when under the magnetic field,
e.g., 
the Landau levels (LLs) in MoS$_{2}$
with a
large band gap grows linearly with the magnetic field,
that's indeed different to the case described in Ref.\cite{Orlita M}
where the linear growth of LLs in zinc-blende semiconductor is only exists in low-magnetic field regime.
Moreover, the time-reversal invariance (TRI) in the bilayer silicene may be broken due to the strong trigonal warping effect which could
suppress the backscattering in Fermi surface\cite{XX} due to the weak antilocalization\cite{McCann E}
and such phenomenon also appears in the monolayer graphene in high energy case.

At low-temperature, the carrier transport and the relaxation rate of the bilayer silicene is discussed
in the presence of randomly charged impurity and the dominant elastic back scattering,
we also 
compare the results to the case of three-dimension Weyl semimetal.
The electronic transport in the presence of screened Coulomb potential also affects the dc conductivity\cite{Hwang E H}
as well as the Friedel oscillation\cite{Wu C H3,XX,Wu C H_3,Wu C H_4,Cheianov V V} of the silicene.
At finite temperature,
the the electron-phonon interaction is calculated and the acoustic phonon model of silicene is obtained by the DFT study 
and compared to the MoS$_{2}$.
For long-range Coulomb potential in the Thomas-Fermi approximation,
the external potential induced by the electric field or the impurity within substrate
is also related to the induced charge density.

This paper is organized as follows.
In Sec.2,
we present the low-energy effective Dirac model of the bilayer silicene.
In Sec.3, we mainly discuss the dynamical polarization and the carrier-density-dependence of the bilayer silicene.
In Sec.4, we discuss the Boltzmann transport in the dirty sample of bilayer silicene in the presence of the long-range impurity.
In Sec.5,
the long-range electron scattering rate at finite temperature in the presence of electron-phonon interaction is obtained by the Fermi's golden rule,
and the intravalley scattering by the acoustic phonon is derived by the deformation-potential approximation,
In Sec.6, we discuss the relaxation rate (in the case of short-range potential)
of the Dirac and Weyl system for a comparation,
The static polarization of bilayer silicene is obtained.
In Sec.7, we discuss the long-range potential.
In the absence of gate voltage, there are not free carriers at zero-temperature\cite{Hwang E H}.
We consider an electric field along transport direction, then the temperature-dependent mobility can be obtained.
In Appendix.A, we detailly describe the low-energy effective Dirac model of the bilayer silicene as well as
its eigenvalue.
In Appendix.B, we presence the detail derive of the dynamical polarization of bilayer silicene
at finite temperature with the strong interlayer hopping.

\section{Model}

For the dynamical polarization of the monolayer silicene, we have discussed in the Refs.\cite{XX,Wu C H3,Wu C H_3,Wu C H_4},
while for the bilayer silicene which is a parabolic system
the interlayer hopping $t_{\perp}$ need to be taken into account.
The kink of the bilayer silicene we discuss here is the AB-stacked one which is generally more stable than the AA-stacked one\cite{Liu F},
and the parameters are setted as:
The nearest-neighbor (NN) interlayer hopping is $t_{\perp}=2$ eV
which is much larger than the next-nearest-neighbor (NNN) interlayer one\cite{Liu F} and thus we consider only the NN interlayer hopping here.
The interlayer spin-orbit coupling (SOC) $\lambda_{SOC\perp}$ is estimated as 0.5 meV here\cite{Ezawa M2} and since the 
trigonal warping term between two layers has a non-negligible impact (expecially when applies the light in terahert range\cite{Morell E S}),
we set the trigonal warping hopping parameter as $t_{w}=0.16$ eV.
Then the low-energy Dirac effective model can be written as:
\begin{equation} 
\begin{aligned}
H=\eta\begin{pmatrix}m_{D}^{\eta ++}&\hbar v_{w}(k_{x}+ik_{y})&0&\hbar v_{F}(k_{x}-ik_{y})\\
 \hbar v_{w}(k_{x}-ik_{y})    &m_{D}^{\eta +-}&\hbar v_{F}(k_{x}+ik_{y})&0\\
 0&\hbar v_{F}(k_{x}-ik_{y})   &m_{D}^{\eta -+}&\eta t'\\        
 \hbar v_{F}(k_{x}+ik_{y})   &0&\eta t'&m_{D}^{\eta --}\\  
 \end{pmatrix},
\end{aligned}
\end{equation}
where $v_{w}=\sqrt{3}at_{w}/2\hbar\sim v_{F}/10\sim 5.5\times 10^{4}m/s$ is the velocity associates with the trigonal warping,
$v_{F}=5.5\times 10^{5}$ m/s is the Fermi velocity of the freestnding silicene and it can be $v_{F}=1.2\sim 1.3\times 10^{6}$ m/s 
in a Ag-substrate.
The trigonal warping term breaks the valley symmetry and will leads to the single-Dirac-cone state
under the light in a finite intensity which is simialr to the effect of the out-of-plane antiferromagnetic exchange field.
The above matrix can also be written as
\cite{Wu C HX,Wu C H3,Wu C H1,Wu C H4,Wu C H7,XX,Wu C H_3,Wu C H5,Ezawa M2}
\begin{equation} 
\begin{aligned}
H=&\hbar v_{F}(\eta\tau_{x}k_{x}+\tau_{y}k_{y})+\eta\lambda_{{\rm SOC}}\tau_{z}\sigma_{z}+a\lambda_{R_{2}}\eta\tau_{z}(k_{y}\sigma_{x}-k_{x}\sigma_{y})\\
&-\frac{\overline{\Delta}}{2}E_{\perp}\tau_{z}+\frac{\lambda_{R_{1}}}{2}(\eta\sigma_{y}\tau_{x}-\sigma_{x}\tau_{y})+M_{s}\sigma_{z}
+\lambda_{SOC\perp}\tau_{z}(\eta\sigma_{x}\tau'_{y}+\sigma_{y}\tau'_{x})\\
&+\frac{t_{\perp}}{2}(\tau_{x}\tau'_{x}-\tau_{y}\tau'_{y})
+\frac{t_{w}}{2}(k_{x}\tau_{x}\tau'_{x}+k_{x}\tau_{y}\tau'_{y})
+\mu,
\end{aligned}
\end{equation}
where 
$E_{\perp}$ is the perpendicularly applied electric field, 
$a=3.86$ is the lattice constant,
$\mu$ is the chemical potential,
$\tau'_{x/y}$ denotes the pseudospin of the layers,
$\overline{\Delta}=0.46$ \AA\ is the buckled distance between the upper sublattice and lower sublattice,
$\sigma_{z}$ and $\tau_{z}$ are the spin and sublattice (pseudospin) degrees of freedom, respectively.
$\eta=\pm 1$ for K and K' valley, respectively.
$M_{s}$ is the spin-dependent exchange field. 
$\lambda_{SOC}=3.9$ meV is the strength of intrinsic spin-orbit coupling (SOC) and $\lambda_{R_{2}}=0.7$ meV is the intrinsic Rashba coupling
which is a next-nearest-neightbor (NNN) hopping term and breaks the lattice inversion symmetry.
$\lambda_{R_{1}}$ is the electric field-induced nearest-neighbor (NN) Rashba coupling which has been found that linear with the applied electric field
in our previous works\cite{Wu C H1}: $\lambda_{R_{1}}=0.012E_{\perp}$.
The Dirac-mass for one spin (or pseudospin) component can be obtained throught the diagonalization procedure:
\begin{equation} 
\begin{aligned}
&m_{D}^{\eta,\sigma_{z},\tau_{z}}=\eta\sqrt{\lambda_{{\rm SOC}}^{2}+a^{2}\lambda^{2}_{R_{2}}k^{2}}\sigma_{z}\tau_{z}-\frac{\overline{\Delta}}{2}E_{\perp}\tau_{z}+M_{s}\sigma_{z},\\
\end{aligned}
\end{equation}
The Dirac-mass is related to the band gap by the relation $\Delta=2|m_{D}|$.
The energy $\varepsilon$ can be obtained by solving the Eq.(2)
as shown in Appendix.A.
A significant difference of bilayer silicene to the monolayer silicene is the trigonal warping due to the 
strong interlayer hopping which is replaced by the hexagonal warping\cite{XX} 
as observed upon the Ag(111) substrate\cite{Feng B}.
The energy and the DOS distribution of the monolayer silicene (with hexagonal warping (HW)) and bilayer silicene
(with trigonal warping (TW)) are presented in the Fig.1,
In Fig.2, we show the band structure of bilayer silicene
under electric field.
The minimum band gap ($m_{D}^{\rm min}$) vanishes in the critical electric field.
In Fig.2, the interlayer hopping also cause a slightly asymmetry of the band structure 
which can be seen when compared to the upper panels (without consider the interlayer hopping).

\section{Fermi wave vector, DOS at Fermi level, and the dynamical polarization of bilayer silicene
in the presence of impurity}

In the presence of the impurities as well as the short- or long-range Coulomb interaction,
an important quantity is the dynamical polarization which estimates the screening static impurity potential
or the screening due to the collective excitation models\cite{Wu C H3,Wu C H_3,XX},
and renormalizes the Coulomb interactions between the carriers.
We have clarified the density-dependence (carriers or the impurities) of the Fermi wave vector, Thomas-Fermi wave vector,
plasmon dispersion,
and the density of states (DOS) at Fermi level in Ref.\cite{Wu C H_4}
for the 2D Dirac system and the 3D Dirac or Weyl system,
here we use the results we obtained in Ref.\cite{Wu C H_4} for further discussion.
As we discussed in Ref.\cite{Wu C H_4},
the Fermi wave vector of monolayer 2D Dirac system is ${\bf k}_{F}^{(2)}=\sqrt{4\pi n/g_{s}g_{v}}$,
where $n$ is the carrier density and $g_{s}g_{v}=4$ denotes the spin and valley degrees of freedom.
And for a typical value of carrier density $n=1\times 10^{12}$ cm$^{-2}$, the Fermi wavevectors can be obtained as
${\bf k}_{F}^{(2)}=\sqrt{n\pi}=1.8\times 10^{8}$ m$^{-1}\approx 5.56$ nm.
While for the bilayer silicene (or other parabolic 2D systems),
due to the existence of $t_{\perp}$,
the Fermi wave vector becomes
${\bf k}_{F}^{bi}=t_{\perp}\sqrt{4\pi n/g_{s}g_{v}}=\sqrt{4\pi \mu(\mu+t_{\perp})/g_{s}g_{v}}$,
i.e., the interlayer hopping has the relation $t_{\perp}=\frac{\mu+\mu\sqrt{4n+1}}{2n}$.
For the maximum carrier density $n=10^{13}$ cm$^{-2}$,
the Fermi wave vector can be obtained as ${\bf k}_{F}=
=1.06\times 10^{7}$ m$^{-1}\approx 94$ nm, where the interlayer hopping here is $t_{\perp}=0.0189$ eV.
We next simplify the maximum carrier density as $n_{{\rm max}}=1$,
then,
for a experimental achievable chemical potential $\mu=0.2$ eV,
through above expression we obtain $t_{\perp}=\frac{\mu+\mu\sqrt{4+1}}{2}=0.323$ eV,
thus $\mu/t_{\perp}=0.2/0.323=0.619$, which is close to the result of Ref.\cite{Gamayun O V} (0.6);
while for the smaller carrier density $n=10^{12}$ cm$^{-2}=n_{{\rm max}}/10$, it becomes
$t_{\perp}=\frac{\mu+\mu\sqrt{4\times 0.1+1}}{2\times 0.1}=1.63$ eV,
then $\mu/t_{\perp}=0.2/1.63=0.12$,
which close to the result of Ref.\cite{Gamayun O V} again (0.2).
Note that here the deviation here may due to the neglect of the factor $\pi$ in the Ref.\cite{Gamayun O V}.
As shown in Ref.\cite{Wu C H_4},
the DOS at Fermi level of the linear 2D Dirac system reads $D_{F}^{(2)}=\sqrt{g_{s}g_{v}\frac{n\pi}{\gamma}}$
where $\gamma$ is the band parameter which can be estimated as $v_{F}$ in the case of large distance.
For the bilayer silicene, the DOS reads $D_{F}^{bi}=g_{s}g_{v}\frac{t_{\perp}+2\mu}{4\pi}$,
and the resulting dielectric function is
\begin{equation} 
\begin{aligned}
\epsilon^{bi}({\bf q},\omega)=\epsilon^{*}(1+\frac{V_{q}D_{F}^{bi}}{\epsilon^{*}\hbar v_{F}}\Pi^{bi}({\bf q},\omega)),
\end{aligned}
\end{equation}
where $V_{q}=\frac{2\pi e^{2}}{\epsilon{\bf q}}$ is the 2D Fourier transform of the Coulomb interaction,
and here $e^{2}/\hbar v_{F}$ can be estimated as 2.16\cite{Wu C H_3,Khveshchenko D V}.
$\epsilon^{*}$ is the effective background dielectric constant\cite{Wu C H_4,Lv M,Zhou J}.
The expression of the 
dielectric function for bilayer silicene is differ from the monolayer one
where the factor $D_{F}^{bi}$ is missing.
The dynamical polarization of the bilayer silicene (or other parabolic chiral system) reads
\begin{equation} 
\begin{aligned}
\Pi^{bi}({\bf q},\omega)=-g_{s}g_{v}\frac{2\pi e^{2}}{\epsilon_{0}\epsilon}\sum_{m_{D}}
\int_{1st BZ}\frac{d^{2}k}{(2\pi)^{2}}\sum_{s,s'=\pm 1}\frac{f_{s,({\bf k}+{\bf q})}-f_{s',{\bf k}}}{E_{s,({\bf k}+{\bf q})}
-E_{s',{\bf k}}-\Omega-i\delta}{\bf F}_{ss'}({\bf k},({\bf k}+{\bf q})),
\end{aligned}
\end{equation}
with
\begin{equation} 
\begin{aligned}
{\bf F}_{ss'}({\bf k},({\bf k}+{\bf q}))
=\frac{1}{2}+\frac{ss'}{2}{\rm cos}(2\theta_{ss'}),
\end{aligned}
\end{equation}
where $\theta_{ss'}={\rm acos}\frac{{\bf k}\cdot{\bf k}'}{kk'}$ is the angle between the wave vectors before and after scattering,
then we have
\begin{equation} 
\begin{aligned}
{\rm cos}(2\theta_{ss'})=
2(\frac{k+q{\rm cos}\phi+m_{D}^{2}}{\sqrt{k^{2}+q^{2}+2kq{\rm cos}\phi}})^{2}-1
\end{aligned}
\end{equation}
where $\phi$ is the angle between ${\bf k}$ and ${\bf q}$,
and here we define $|{\bf k}+{\bf q}|=\sqrt{k^{2}+q^{2}+2kq{\rm cos}\theta}$.
The area of the unit 2D cell $A$ is setted as 1 for simplicity.
The eigenvalue about the energy $\varepsilon$ can be analytically obtained by solving the Eq.(2). 
Then the dynamical polarization for a low-temperature can be obtained after some algebra.
We 
set $s=1$ which stands for the conduction band,
then at zero temperature,
the imaginary part of the dyamical polarization reads
\begin{equation} 
\begin{aligned}
{\rm Im}\Pi^{bi}({\bf q},\omega)=&-g_{s}g_{v}\frac{ e^{2}}{2\epsilon_{0}\epsilon}
\int^{\sqrt{\Lambda^{2}-m_{D}^{2}}}_{0}kdk
\left\{\frac{1}{4k^{2}}\left[\phi(4m_{D}^{2}k+3k^{2}-q^{2})\right.\right.\\
&\left.\left.+\frac{2(2m_{D}^{2}k+k^{2}-q^{2})^{2}{\rm atan}\frac{{\rm tan}\frac{\phi}{2}(k-q)}{k+q}}{k^{2}-q^{2}}+2kq{\rm sin}\phi\right]\right\}\Bigg|_{\phi}\delta(\omega+i\eta+E_{s,{\bf k}}+E_{s',({\bf k}+{\bf q})})
\end{aligned}
\end{equation}
for $s'=1$;
\begin{equation}
\begin{aligned}
{\rm Im}\Pi^{bi}({\bf q},\omega)=&-g_{s}g_{v}\frac{ e^{2}}{2\epsilon_{0}\epsilon}
\int^{\sqrt{\Lambda^{2}-m_{D}^{2}}}_{0}kdk
\left\{\frac{1}{4k^{2}}\left[\phi(-4m_{D}^{2}k+k^{2}+q^{2})\right.\right.\\
&\left.\left.-\frac{2(2m_{D}^{2}k+k^{2}-q^{2})^{2}{\rm atan}\frac{{\rm tan}\frac{\phi}{2}(k-q)}{k+q}}{k^{2}-q^{2}}-2kq{\rm sin}\phi\right]\right\}\Bigg|_{\phi}\delta(\omega+i\eta+E_{s,{\bf k}}-E_{s',({\bf k}+{\bf q})})
\end{aligned}
\end{equation}
for $s'=-1$.
Here $\Lambda\gg m_{D}+q$ as a integration limit.
While for the case of $s=-1$,
the dynamical polarization can be deduced by the same way.
The term $\delta(\omega+i\eta+E_{s,{\bf k}}-E_{s',({\bf k}+{\bf q})})$ in above expressions
also appear in the Kubo-Greenwood formula of the mobility\cite{Fischetti M V}.
The real part of the polarization can be obtained through the Kramers-Kronig relation
\begin{equation} 
\begin{aligned}
{\rm Re}[\Pi({\bf q},\Omega)]
=\frac{1}{\pi}\mathcal{P}\int^{\infty}_{-\infty}d\omega\frac{\omega{\rm Im}[\Pi({\bf q},\omega)]}{\omega^{2}-\Omega^{2}}{\rm sgn}[\omega].
\end{aligned}
\end{equation}
At zero-temperature,
the Fermi-Dirac distribution function can be replaced by the step function,
$f_{1,({\bf k}+{\bf q})}=\theta(k_{F}^{bi}-k-q)=\theta(\frac{\sqrt{2m^{*}\mu}}{\hbar}-k-q)$, $f_{-1,({\bf k}+{\bf q})}=1$.
Since elastic backscattering is dominant for both the 2DEG and the bilayer silicene in low-temparature\cite{Adam S},
the angle $\phi$ can be setted as $\pi$,
then
the result of the chirality factor ${\bf F}_{ss'}$ for $s'=1$ (intraband) and $s'=-1$ (interband) are shown in Fig.3,
where the peak appear in ${\bf q}={\bf k}_{F}^{bi}\approx 0.327$ eV.
Here we note that, althrough the 2DEG and the bilayer silicene have many similarities in electronic transport properties
as well as the carriers density-dependence during the self-consistent transport\cite{Wu C H3,Sensarma R},
their polarization function have different forms:
The polarization of the 2DEG does not has the chiral factor ${\bf F}_{ss'}$,
and thus its relaxation time in a form as shown in Ref.\cite{Jena D}.
The missing of the chirality in 2DEG also gives rise the backscattering unlike the silicene or the Weyl semimetals.

\section{Boltzmann transport and the density-dependence in bilayer silicene}

Next we analyze the effect of the impurity to the electron transport.
In the presence of long-range impurity (not the $\delta$-type), 
we first assuming the collision-free Boltzmann equation in relaxation time approximation:
\begin{equation} 
\begin{aligned}
\dot{f_{k}}+\nabla_{r}\cdot(v_{k}f_{k})=-\frac{f_{k}-f_{k}^{0}}{\tau_{k}},
\end{aligned}
\end{equation}
where $v_{k}$ is the momentum-dependent group velocity.
Here the external field-induced term $\nabla_{k}\cdot f_{k}$ is missing and thus guarantee the isotropic scattering.
Then in the Hartree-Fock approximation for the on-site interaction and the self-consistent Born approximation 
in the presence of impurity with the disorder scattering potential in density $n_{s}=n_{i}^{{\rm scatt}}$
(in 2D 
electron system) where $n_{i}$ is the impurity concentration.
The $n_{s}$ here
plays an important role in room temperature mobility\cite{Shen L}:
The lower the disorder scattering, the higher electron mobility.
The collision rate 
in the presence of the impurity scattering reads
\begin{equation} 
\begin{aligned}
\frac{1}{\tau}=
\frac{n_{s}}{h\hbar v_{F}}\frac{\varepsilon_{F}}{\hbar v_{F}}\int^{2\pi}_{0} d\theta
\frac{1+{\rm cos}\theta+\delta(E_{k}-E_{k'})(1-{\rm cos}\theta)}{2}|U({\bf q})|^{2}
(1-{\rm cos}\theta),
\end{aligned}
\end{equation}
where $\theta$ is scattering angle.
$U({\bf q})=\frac{e^{2}}{2\epsilon\epsilon^{*}\sqrt{{\bf q}^{2}+{\bf k}_{s}^{2}}}$ is the 2D Fourier transform of the 
$U(r)=\frac{e^{2}e^{-{\bf k}_{s}}}{4\pi\epsilon\epsilon^{*}}$.
The relaxation time is inverse proportional to the conductivity since it's  proportion to the impurity concentration $n_{i}$.
The scattering wave vector reads
${\bf q}={\bf k}-{\bf k}'=2k{\rm sin}(\theta/2)$
and here $k\sim \frac{\varepsilon}{\hbar v_{F}}\sim n_{i}^{1/2}$ for the 2D Dirac system.
The standard\cite{Burkov A A,Nomura K} factor $(1-{\rm cos}\theta)$ exist in the presence of the dominating elastic backscattering
and isotropic relaxation for the 2D or 3D Dirac or Weyl system,
but for bilayer system, it becomes $(1-{\rm cos}2\theta)$ due to the special pseudospin distribution.
The $\delta$-term here can be preserved even for the long-range impurity as a result of low-temperature,
and the $\delta$-term also proportional to the strength of the intraband transition.
In the low-temperature case, the relaxation process is affected by the (scattering) phase-space restriction\cite{phase}
with the Pauli principle\cite{Pines D},
in addition, we have ${\rm sin}\theta=\hbar v_{F}{\bf k}/\varepsilon$ if the angle $\theta$ is in the phase-space.
While in Eq.(5), we does not consider the chirality and the pseudospin distribution in phase-space.
In the zero-field limit, the dc conductivity reads $\sigma_{0}=\frac{2e^{2}}{h\hbar}E_{F}^{(2)}\tau$ with the Fermi energy
$E_{F}^{(2)}=\gamma {\bf k}_{F}\approx v_{F} {\bf k}_{F}$.
Since in the gapless case, the $\frac{1}{\tau}$ reads
\begin{equation} 
\begin{aligned}
\frac{1}{\tau}=&
\frac{n_{s}}{h\hbar v_{F}}{\bf k}_{F}\int^{2\pi}_{0} d\theta\frac{1}{2}|U({\bf q})|^{2}
(1-{\rm cos}^{2}\theta)\\
=&\frac{n_{s}}{h\hbar v_{F}}{\bf k}_{F}\frac{\pi}{2}|U({\bf q})|^{2},
\end{aligned}
\end{equation}
the dc conductivity $\sigma_{0}$ becomes independent of the Fermi wavevector,
and we can also know that, for the case of elastic backscattering

The short-range impurity scattering is dominating in the absence of electron interaction.
With the screened Coulomb interaction in the presence of a large experimentally accessible carrier density, 
e.g., in a density $n=4.35\times 10^{12}$ cm$^{-2}$, the Thomas-Fermi wave vector can be evaluated 
by the self-consistent Thomas-Fermi screening approximation (${\bf q}_{TF}>2{\bf k}_{F}$)\cite{Resta R}
as ${\bf q}_{TF}=g_{s}g_{v}r_{s}{\bf k}_{F}=1.183\times 10^{7}$ cm$^{-1}$ 
(or the inverse screening length) where 
$r_{s}=\frac{e^{2}}{\hbar \epsilon\epsilon^{*}\gamma}$ is the Wigner-Seitz radiu.
While for the parabolic system, the Thomas-Fermi wave vector is independent of the carrier density
but relys more on the interlayer hopping,
which reads
${\bf q}_{TF}^{bi}=g_{s}g_{v}r^{bi}_{s}{\bf k}_{F}=g_{s}g_{v}\frac{e^{2}m^{*}}{{\bf k}_{F}\epsilon\epsilon^{*}\hbar^{2}}{\bf k}_{F}
=g_{s}g_{v}\frac{e^{2}m^{*}}{\epsilon\epsilon^{*}\hbar^{2}}$,
with $r^{bi}_{s}=\frac{e^{2}m^{*}}{{\bf k}_{F}\epsilon\epsilon^{*}}$
the effective mass in silicene $m^{*}=t_{\perp}/(2 v_{F}^{2})=2\hbar^{2}t_{\perp}/(3 a^{2}t^{2})\sim 1/v_{F}^{2}$
(using the identity $\hbar v_{F}=\frac{\sqrt{3}}{2}at$).
For bilayer silicene, $m^{*}=0.268 m_{e}$, and it is larger than the bilayer graphene which is obtained as 0.033 $m_{e}$\cite{Adam S},
then we can obtain ${\bf q}_{TF}^{bi}\approx 8.124\times 10^{9}$ m$^{-1}$
which is larger than the one of bilayer graphene (about $8\times 10^{9}$ m$^{-1}$).
We note here that, although the Thomas-Fermi wave vector in bilayer silicene is independent of the carrier density,
it restore the carrier-density-dependence in
the 3D Dirac or Weyl system (i.e., the bulk form of the host materials),
which becomes ${\bf q}_{TF}=3\times \frac{g_{s}g_{v}\pi n e^{2}}{\epsilon\epsilon^{*}E_{F}}$.
For the Donor impurity in 3D Weyl semimetal, we also have\cite{Burkov A A} ${\bf q}_{TF}^{2}=4\pi e^{2}(D_{F}^{(3)})^{3}$
where $D_{F}^{(3)}=\sqrt[3]{9g_{s}g_{v}n^{2}/2\pi^{2}\gamma^{3}}\ll 1$ is the DOS of Fermi level in 3D Weyl semimetal
with the long-range Coulomb potential.

For the overscreened Coulomb intercation (${\bf q}_{TF}\gg 2{\bf k}_{F}$\cite{Adam S}),
both the 2DEG and bilayer silicene have a dc conductivity $\sigma\propto n/n_{i}$\cite{Nomura K2,Adam S}.
In coherent phase approximation,
the mean-free path $\ell=v_{F}\tau$ is related to the $D_{F}$
by $D_{F}\propto 1/\ell v_{F}$.
In addition, the self-doping effect as well as the value of ${\bf k}_{F}\ell$ will enhanced with increase of $n_{i}$,
e.g., for $n_{i}=10^{12}$ cm$^{-2}$, ${\bf k}_{F}=5.64\times 10^{-9}$ m, thus ${\bf k}_{F}\ell=0.1$,
while for $n_{i}=10^{10}$ cm$^{-2}$, ${\bf k}_{F}\ell=1$\cite{Stauber T}.

In low-energy limit ($\varepsilon<0.1$ eV) where the parabolic structure of the bilayer silicene is broken down,
the residual carrier density exists in Dirac-point with the linear band structure.
We note that the lower residual carrier density in silicene results in a smaller bandgap compared to the graphene\cite{Tao L}.
The residual carrier density is the carrier density (the electron and hole) in Dirac-point
given by a sum of the populated states\cite{Tao L} including the ones induced by the charged impurities,
and it's exists even in the presence of the broken Dirac-cone by the Ag-substrate\cite{Wang Y P}.
For $\varepsilon>E_{F}$,
we have
\begin{equation} 
\begin{aligned}
n_{r}=&2\int^{\Lambda}_{m_{D}}D(\varepsilon)f(\varepsilon)d\varepsilon\\
=&\frac{4}{\pi\hbar^{2}v_{F}^{2}}\left[\frac{1}{2}\varepsilon(\varepsilon-2T{\rm ln}(e^{(\varepsilon-E_{F})/T}+1))
-T^{2}{\rm Li}_{2}(-e^{(\varepsilon-E_{F})/T})\right]\bigg|^{3t}_{m_{D}},\ \varepsilon>E_{F}
\end{aligned}
\end{equation}
where $f(\varepsilon)=1/(1+e^{(\varepsilon-E_{F})/T})$ is the Fermi-distribution function,
$D(\varepsilon)=\frac{2\varepsilon}{\pi\hbar^{2}v_{F}^{2}}$ is the DOS where we consider the two spin flavors here.
${\rm Li}_{2}(x)$ is the dilogarithm function.
$\Lambda$ is the high-energy cutoff which can be estimated as the bandwidth $W=3t$ here.
The plot of $n_{r}$ is shown in Fig.4,
where the residual density seems increase monotonically as temperature increase
and nearly independent of the chemical potential.
For $\varepsilon=E_{F}$, we have $n_{r}=D_{F}\propto \sqrt{n_{i}}$.
The influence of the impurity concentration to the residual carrier density can be ignored only in the case that $2|m_{D}|\ll k_{B}T$,
i.e., the band gap is smaller than the thermal activation,
and thus the effect of the impurity is important in the low-temperature case as shown in the low-energy Dirac system.
The dc minimal conductivity is proportional to the carrier density as 
$\sigma_{dc}^{{\rm min}}=\frac{e^{2}v_{F}}{h}\tau{\rm max}[{\bf k}_{F},\pi\alpha\sqrt{n}]$\cite{Stauber T}
where $\alpha\leq 1$ is a dimensionless constant.
The minimal dc condictivity induced by static gate voltage reads $\sigma_{dc}=20\frac{e^{2}}{h}\frac{{\rm max}[n,n_{r}]}{n_{i}}$\cite{Adam S2}
for the SiO$_{2}$ subatrate which with the Wigner-Seitz radiu $r_{s}\approx 0.553$.
The general relation $\sigma\sim n/n_{i}$ is been proved\cite{Adam S} that valid for the case of screened Coulomb potential.
For small impurity concentration $n_{i}=5\times 10^{11}$ cm$^{-2}$ in SiO$_{2}$, the corresponding $\sigma_{dc}=7.6\frac{e^{2}}{h}$
with $n/n_{i}=0.38$.
In diffusive kinetic theory, the conductivity has $\sigma\sim {\bf k}_{F}\ell\sim \rho^{-1}\sim \tau^{2}\sim 1/n_{i}$,
where $\rho$ is the resistivity,
the scattering time $\tau$ increase with the carrier density, and for high mobility which corresponds to short-range scattering,
the rate of increase of $\tau$ with $n$ is lower than the low-mobility one.

In the presence of the disorder to the silicene lattice system,
it's found that not localized state exist in the massless silicene\cite{Nomura K3}.
For monolayer silicene and bilayer silicene,
the Berry phases accumulated near Dirac-cone are $\pi$ and $2\pi$, respectively, similar to the gapless graphene,
and they have a destructive interference effect between the backscattering and its time-reversal counterpart behavior\cite{Nomura K3,Ando T}.
Thus the backscattering is supressed by the time-reversal in this case
unless has stronger disorder to breaks the chirality
(or when the disorder dominates over the effect of interactions at some
fixed point of the momentum space).
The supression of the backscattering also results from of the presence of quasiparticle chirality 
no matter in the 3D Weyl system\cite{Lv M} or 2D Dirac system\cite{XX,Wu C H_3,Wu C H_4}.


\section{Phonon}

In finite temperature (or up to room temperature) where the Fermi-Dirac distribution function within the expression of $n_{r}$ describes the 
occupation of the phonon, the Monte Carlo simulation is valid during the calculation of the phonon transport
in full-band analysis\cite{Li X2}.
Here we present in Table.1 the first principle result of the (acoustic) phonon model of the silicene and MoS$_{2}$ by
using the QUANTUM ESPRESSO package\cite{Paolo} where
the plane wave energy cutoff is setted as 400 eV for the ultrasoft pseudopotential
and the structures are relaxed until the Hellmann-Feynman force on each atom is below 0.01 eV/\AA\ .
The phonon frequency here is also helpful in determining the superconductivity transition temperature\cite{Wan W}.
Our result for MoS$_{2}$ is in great agreement with Ref.\cite{Li X2}.
As it's well known, 
the optical phonon spectrum is prevail over the acoustic phonon spectrum,
and
the longitudinal acoustic branch (LA) is usually higher than the 
transverse acoustic branch (TA) which can be easily obtained by analyzing the atomic vibrations.
While for the out-of-plane acoustic branch ($ZA$),
although it has a lowest phonon energy (and thus has a highest phonon scattering rate or electron-phonon coupling constant\cite{Wan W}),
it's generally not to be ignored due to the broken inversal symmetry in silicene, unlike in graphene.
Similarly, the longitudinal strain has a larger effect on the $\pi$-band structure or
the thermal relaxation than the torsional or transverse strain\cite{Pennington G}.

We also want to note that the acoustic phonon models for silicene contribute mainly to the intervalley scattering
(for short-range scattering),
unlike the optical phonon models which contribute to both the intravalley and intervalley scattering.
Through the data shown in Table.1, we know that the phonon energy of monolayer MoS$_{2}$ is higher than the silicene,
and thus the MoS$_{2}$ should has lower scattering rate than silicene,
which can be explained as below.
The long-range electron scattering rate then can be obtained as\cite{Borysenko K M,Li X2,Wan W}
\begin{equation} 
\begin{aligned}
\frac{1}{\tau_{ph}}=\frac{2\pi}{\hbar}\sum_{{\bf q},\nu}|g_{\nu}|^{2}[f_{\nu}\delta(\varepsilon_{{\bf k}+{\bf q}}-\hbar \omega_{\nu}-\varepsilon_{{\bf k}})
+(f_{\nu}+1)\delta(\varepsilon_{{\bf k}-{\bf q}}+\hbar \omega_{\nu}-\varepsilon_{{\bf k}})](1-{\rm cos}\theta),
\end{aligned}
\end{equation}
where $\nu=LA,\ TA,\ ZA$ is the indices of the acoustic branches, 
$g_{\nu}$ is the electron-phonon interaction matrix element\cite{Baroni S}
which scaled by the atomic mass and phonon frequency as
$g_{\nu}=\sqrt{\frac{\hbar}{2m\omega_{\nu}}}\langle \psi_{{\bf k}'}|\delta V_{{\bf q}}|\psi_{{\bf k}}\rangle$
where $\psi$ is the Bloch wave function in reciprocal lattice vector space.
$\delta V_{{\bf q}}$ is the derivative of the self-consistent potential
$V_{{\bf q}}=\sum_{{\bf q}}v_{{\bf q}}({\bf r})e^{i{\bf q}\cdot{\bf r}}$
(we assume the static case here)
with $v_{{\bf q}}({\bf r})$ the lattice-periodic potential\cite{Baroni S}.
The $g_{\nu}$ can be obtained by the density functional perturbation theory (DFPT)
for the periodic potential with finite ${\bf q}$.
Note that the optical limit can't be applied in the DFPT
and thus the macroscopic dielectric matrix can't be approximated by the head of the microscopic dielectric matrix\cite{Gajdo? M},
i.e., preserved the local field (off-diagonal) terms.
$f_{\nu}=1/(e^{\hbar\omega_{\nu}/T}+1)$ describes the phonon occupation, the corresponding phonon frequency $\omega_{\nu}$ can be found in Table.1.
Note that this scattering rate obtained by Fermi golden rule is valid for long-range scattering,
and incorparates the long-range divergence of the Coulomb interaction in the Dirac-point,
but it's unvalid for the short-range scattering (e.g., when the potential range is shorter than the lattice constant $a=$3.86 \AA\ ).
The scattering rate of phonon (10$^{13}$ s$^{-1}$) is much larger than of the impurity (10$^{10}$ s$^{-1}$).
Here we note that for the massless Dirac system in all dimensions,
the dispersion obeys $\varepsilon=v_{F}{\bf k}$,
and the phonon frequency in Debye approximation also has $\omega_{p}=v_{s}{\bf k}$.
For the intravalley scattering of silicene,
and in the presence of broken sublattice symmetry by the electric field, light field, or the tensile strain,
the acoustic phonon scattering rate for single flavor is $1/\tau_{ph}=\frac{D^{2}T\hbar v_{F}{\bf k}_{F}}{\hbar^{3}v_{\nu}^{2}}\rho v_{F}^{2}$
\cite{Li X2,Stauber T,Borysenko K M}
where $\rho=7.2\times 10^{-8}$ g/cm$^{2}$ is the mass density,
$v_{\nu}\approx 2\times 10^{4}$ m/s$\sim 0.036 v_{F}$ is the sound velocity,
$D$ is the electron acoustic deformation potential.
Since the intervalley scattering is dominant in the case of short-range potential,
this intravalley acoustic phonon scattering is much weaker than the intervalley one;
however, the intravalley optical phonon scattering is comparable with the intervalley one\cite{Li X2}.
For the case of electron-hole symmetry in gapless case,
the full relaxation rate (contains all of the flavors) can be written as
$1/\tau_{ph}=\frac{D^{2}T\hbar v_{F}{\bf k}_{F}}{8\hbar^{3}v_{\nu}^{2}}\rho v_{F}^{2}$\cite{Stauber T}.

\section{Relaxation rate and static polarization}

For the short-range (contact potential) scattering in Born approximation,
where the intervalley scattering becomes dominating and the impurity potential can be replaced by a $\delta$-function,
the relaxation rate in 2D Dirac system (monolayer or bilayer) and 3D Weyl system read
\begin{equation} 
\begin{aligned}
\frac{1}{\tau^{(2)}}=\frac{n_{i}V^{(2)2}({\bf q})}{\hbar^{2}v_{F}^{2}}|E_{F}^{(2)}|
=\frac{n_{i}V^{(2)2}({\bf q})}{\hbar^{2}v_{F}^{2}}|\hbar v_{F}{\bf k}_{F}^{(2)}|,\\
\frac{1}{\tau^{(3)}}=\frac{n_{i}V^{(3)2}({\bf q})}{\pi\hbar^{3}v_{F}^{3}}|E_{F}^{(3)}|^{2}
=\frac{n_{i}V^{(3)2}({\bf q})}{\hbar^{2}v_{F}^{2}}\left|\frac{\hbar^{2}({\bf k}_{F}^{(3)})^{2}}{2m^{*}}\right|^{2},\\
\end{aligned}
\end{equation}
where
the 2D Fourier transform of the screened Coulomb interaction is $V^{(2)}({\bf q})=\frac{2\pi e^{2}}
{\epsilon_{0}\epsilon\sqrt{{\bf q}^{2}+{\bf k}_{s}^{2}}}$,
with the screening wave vector 
(or the inverse screening length) 
in static-limit ${\bf k}_{s}=2\pi e^{2}\Pi({\bf q},0,T)/(\epsilon_{0}\epsilon)$ which is polarization-dependent.
While for bilayer silicene, the screened Coulomb interaction is
$V^{bi}({\bf q})=\frac{2\pi e^{2}e^{-{\bf q}|z|}}
{\epsilon_{0}\epsilon\sqrt{{\bf q}^{2}+{\bf k}_{s}^{2}}}$
where $z$ is the distance between the upper (or lower) layer to the middle point of the bilayer silicene
(the short-range impurity is placed in the silicene sheet but not in the substrate now),
and the relaxation rate can be obtained by replacing the $E_{F}^{(2)}$ within $1/\tau^{(2)}$ as $E_{F}^{bi}=\varepsilon$
(see Appendix.A).
For 3D Dirac or Weyl system,
$V^{(3)}({\bf q})=\frac{4\pi e^{2}}
{\epsilon_{0}\epsilon({\bf q}^{2}+{\bf k}_{s}^{2})}$.
The 2D Fermi wave vector ${\bf k}_{F}^{(2)}$ has been presented in above, while
for 3D Weyl system, ${\bf k}_{F}^{(3)}=\sqrt[3]{6\pi^{2}n/g_{s}g_{v}}$
where we assume that the two Weyl nodes are equally populated,
and for bilayer system ${\bf k}^{bi}_{F}=\frac{\sqrt{2m^{*}\mu}}{\hbar}$.
Specifically, for bilayer system in long-wavelength limit at zero temperature,
we have $\Pi^{bi}({\bf q},0)\approx D_{F}^{bi}$
in the case $\mu\rightarrow 0$\cite{Low T},
thus we have ${\bf k}_{s}\approx 2\pi e^{2}D_{F}^{bi}/(\epsilon_{0}\epsilon)$ for bilayer system.

The static polarization in zero temperature $\Pi({\bf q},0,0)$ for 2D Dirac system reads
\cite{Wu C H3,XX,Wu C H_3,Tabert C J,Wu C H_4}
\begin{equation} 
\begin{aligned}
\Pi({\bf q},0,0)=-g_{s}g_{v}\frac{2e^{2}\mu}{\epsilon\epsilon^{*}\hbar^{2}v_{F}^{2}}
\left[\frac{ m_{D}}{2\mu}+\frac{\hbar^{2} v_{F}^{2}{\bf q}^{2}-4 m_{D}^{2}}{4\hbar v_{F}{\bf q}\mu}{\rm asin}\sqrt{\frac{\hbar^{2}v_{F}^{2}{\bf q}^{2}}{\hbar^{2}v_{F}^{2}{\bf q}^{2}+4 m_{D}^{2}}}\right]
\end{aligned}
\end{equation}
for $0<\mu< m_{D}$ (intrinsic), and
\begin{equation} 
\begin{aligned}
\Pi({\bf q},0,0)=-g_{s}g_{v}\frac{2e^{2}\mu}{\epsilon_{0}\epsilon \hbar^{2} v_{F}^{2}}
&\left[1-\Theta({\bf q}-2{\bf k}_{F})\right.\\
&\left.\times\left(\frac{\hbar^{2} v_{F}^{2}\sqrt{{\bf q}^{2}-4{\bf k}_{F}^{2}}}{2\hbar v_{F}{\bf q}}-\frac{\hbar^{2}v_{F}^{2}{\bf q}^{2}-4 m_{D}^{2}}{4\mu \hbar v_{F}{\bf q}}{\rm atan}\frac{\hbar v_{F}\sqrt{{\bf q}^{2}-4{\bf k}^{2}_{F}}}{2\mu}\right)\right]
\end{aligned}
\end{equation}
for $\mu> m_{D}$ (extrinsic).
For bilayer system\cite{Low T},
the intraband polarization (which dominant here) reads\cite{Kittel C,Low T}
\begin{equation} 
\begin{aligned}
\Pi^{bi}({\bf q},0,0)=-g_{s}g_{v}\frac{m^{*}}{\pi^{2}\hbar^{2}}\int^{{\bf k}^{bi}_{F}}_{0}dp \int^{2\pi}_{0}d\theta\frac{p}{s^{2}-4p^{2}{\rm cos}\theta},
\end{aligned}
\end{equation}
where $p=\sqrt{\frac{m^{*}}{m}}|{\bf k}|$, $s=\sqrt{\frac{m^{*}}{m}}|{\bf q}|$.
While for the case of strong interband transition, the parabolic structure will be broken and acts like the linear structure.
In the presence of finite temperature\cite{Liu Y2},
\begin{equation} 
\begin{aligned}
\Pi({\bf q},0,T)
=\int^{\infty}_{0}d\mu\frac{\Pi({\bf q},0,0)}{4T{\rm cosh}^{2}(\frac{E_{F}(T)-E_{F}(0)}{2T})},
\end{aligned}
\end{equation}
with the Fermi energy $E_{F}^{(2)}(0)=g_{s}g_{v}v_{F}{\bf k}_{F}$
and $E_{F}^{(2)}(T)=g_{s}g_{v}T{\rm ln}({\rm exp}(\frac{n}{T}\frac{\pi\hbar^{2}v_{F}^{2}}{2\varepsilon})-1)$ for the monolayer silicene;
or $E_{F}^{bi}(0)=\frac{\hbar^{2}\pi n}{m^{*}}$
and $E_{F}^{bi}(T)=g_{s}g_{v}T{\rm ln}({\rm exp}(\frac{n\hbar^{2}\pi}{Tm^{*}})-1)$ for the bilayer silicene.
The DOS at Fermi level is incorporated in the expression of the Fermi energy, 
which is, for bilayer system, $D_{F}^{bi}=\frac{g_{s}m^{*}}{2\pi\hbar^{2}}$,
and note that here the Fermi level lies in the bottom of the conduction parabolic band.
In Fig.5 we show the relaxation rate of the impurity scattering in static-limit
and
in the absence or presence of the screening wave vector ${\bf k}_{s}$.
We see that for 2D monolayer Dirac system the relaxation rate is in an order of 10$^{10}$ s$^{-1}$,
while for 3D Weyl system the relaxation rate is in an order of 10$^{20}$ s$^{-1}$.
That's agree with the results of the relaxation time observed:
For the freestanding silicene, the relaxation time is nearly 18.2 ps
while for Weyl semimetal the relaxation time is 25 fs (for the typical material Eu$_{2}$Ir$_{2}$O$_{7}$\cite{Wu C H_4}).
Our results also agree with the ones reported in Ref.\cite{Das P}.

\section{Long-range potential}

For long-range impurity (with the potential range larger than the lattice constant $a$) within the SiO$_{2}$ substrate,
the Thomas-Fermi approximation is valid for estimation of the induced charge density,
which reads $\delta n_{k}=-e^{2}\varphi D_{F}^{bi}$\cite{Stauber T}
where $\varphi$ is the scalar potential induced by the impurity or the external electric field.
For the case of charge inhomogeneous in silicene which may due to the bound state,
the disorder leads to a peak of the DOS in Dirac-point, as directly shown by the DFT study in Ref.\cite{Wu C H1}
where we found that both the impurity and the Hubbard interaction will give rise to the DOS in Dirac-point and 
thus lead to compressibility as well as the instability\cite{Guinea F}.
In the presence of electric field applied along the transport direction
which contributes to the nonequilibrium distribution,
the Boltzmann equation in relaxation time approximation becomes
\begin{equation} 
\begin{aligned}
\dot{f_{k}}+\nabla_{r}\cdot(v_{k}f_{k})-\frac{eE_{x}}{\hbar}\nabla_{k}\cdot f_{k}{\rm cos}\theta=-\frac{f_{k}-f_{k}^{0}}{\tau_{k}},
\end{aligned}
\end{equation}
then carrier mobility (hole)
has $\xi_{x}\sim v_{x}/E_{x}=\frac{\partial H}{\hbar\partial k}\frac{1}{E_{x}}$ where $v_{x}=v{\rm sin}\ \theta$ is the group velocity
in transport direction ($x$-direction here).
The carrier mobility at finite temperature reads\cite{Fischetti M V}
\begin{equation} 
\begin{aligned}
\xi_{x}=\frac{g_{s}g_{v}e}{4\pi^{2}\hbar^{2}n_{h}}\int^{2\pi}_{0}d\phi\int^{\Lambda}_{E_{F}^{bi}}d\varepsilon
\frac{k(E_{x},\phi)}{|v(\phi)|}v_{x}^{2}(\phi)\tau(E_{x},\phi)(-\frac{\partial f_{k}^{0}}{\partial \varepsilon}),
\end{aligned}
\end{equation}
where 
\begin{equation} 
\begin{aligned}
-\frac{\partial f_{k}^{0}}{\partial \varepsilon}=\frac{f^{0}_{k}(1-f^{0}_{k})}{T}
=\frac{1}{g_{s}g_{v}}\frac{q^{2}E_{F}^{bi}}{\pi\sum_{k}({\bf q}\cdot{v_{k}})^{2}}
\end{aligned}
\end{equation}
in relaxation time approximation,
and the hole density here reads $n_{h}=g_{s}g_{v}\sum_{k}\int\frac{d^{2}q}{(2\pi)^{2}}f_{k}(q)
=g_{s}g_{v}\int^{\Lambda}_{0}d\varepsilon f(\varepsilon)D_{F}^{bi}$.

In long-wavelength limit (in the absence of Dirac-quasiparticle scattering),
the static polarization (within the band gap) can be represented by the thermally averaged DOS:
\begin{equation} 
\begin{aligned}
\Pi({\bf q}\rightarrow 0,0)=&\int^{\Lambda}_{0}d\varepsilon D_{F}^{bi}(\varepsilon)\frac{\partial f_{k}}{\partial \varepsilon}\\
=&-\int^{\mu}d\varepsilon \frac{g_{s}g_{v}m^{*}}{2\pi\hbar^{2}}\sqrt{\frac{m_{D}}{\varepsilon}}
\frac{1}{g_{s}g_{v}}\frac{q^{2}E_{F}^{bi}(T)}{\pi\sum_{k}({\bf q}\cdot{v_{k}})^{2}}\\
=&\frac{m^{*}q^{2}\varepsilon}{\pi^{2}\sum_{k}({\bf q}\cdot{\bf v}_{k})^{2}}\sqrt{\frac{m_{D}}{\varepsilon}}E^{bi}_{F}(T)(\varepsilon\le m_{D}),
\end{aligned}
\end{equation}
where the velocity can be obtained as
\begin{equation} 
\begin{aligned}
{\bf v}_{k}=\frac{\partial H}{\hbar\partial k}
=\frac{1}{\hbar}(\hbar v_{F}+a\eta\lambda_{R_{2}}+\frac{t_{w}}{2}),
\end{aligned}
\end{equation}
for a single component of the degree of freedom.
This is in agree with our previous result\cite{Wu C H4} (not in relaxation time approximation):
\begin{equation} 
\begin{aligned}
\Pi(0,0)=-g_{s}g_{v}\frac{2e^{2}T}{2\pi\epsilon_{0}\epsilon v_{F}^{2}}\left[{\rm ln}(2{\rm cosh}\frac{m_{D}+\mu}{T})-\frac{m_{D}}{2T}{\rm tanh}\frac{m_{D}+\mu}{2T}+(\mu\rightarrow -\mu)\right].
\end{aligned}
\end{equation}
at finite temperature, and
\begin{equation} 
\begin{aligned}
\Pi(0,0)_{T\rightarrow 0}=
-e^{2}D(|\mu|)=-e^{2}\frac{g_{s}g_{v}|\mu|}{2\pi\hbar^{2}v_{F}^{2}}\frac{1}{2}\sum_{\eta=\pm 1}\left[\theta(|2\mu|-2|m_{D}|_{\eta})\right].
\end{aligned}
\end{equation}
at zero-temperature.
In Fig.6, we show the static polarization function described above with different temperature and Dirac-mass.
In the upper panel of Fig.6, we show the trend of the static polarization in relaxation-time approximation (Eq.(24))
with the increasing temperature,
we find that for higher temperature, the required carrier density for nonzero polarization is larger.
And with the increase of $T$, the decrease of polarization becomes slower (see lower panel of the Fig.6),
and its temperature-dependence is decrease with the increase of Dirac-mass,
thus we can expect that the polarization could be nearly $T$-independent in adiabatically case (with large gap)
and at high temperature.

The usual handling method for the impurity problem is by assuming the single-impurity case that the imputiry potential scattering is 
given by a $\delta$-term and independent of the momentum.
Such assuming also provides the
electron-density-deviation induced by the impurity reads $\delta n_{k}=\int\frac{d^{2}q}{(2\pi)^{2}}e^{i{\bf q}\cdot{\bf r}}(1-\epsilon({\bf q}))$
(in static limit),
where ${\bf r}$ here denotes the position of the single-impurity, and the term $e^{i{\bf q}\cdot{\bf r}}$ can be omitted when we set ${\bf r}=0$,
i.e., embed the impurity in the position ${\bf r}=0$.
While for the multi-impurity case,
in the method of $T$-matrix approximation\cite{Onoda S},
the $T$-matrix is no longer only depends on the $\omega$,
but also depends on the ${\bf q}$,
i.e., the $T$ satisfies
\begin{equation} 
\begin{aligned}
T(k,k',\varepsilon)=V(k,k')+\sum_{k'=k+q}V(q,q'')G_{0}(q'',\varepsilon)T(q'',q',\varepsilon),
\end{aligned}
\end{equation}
where the non-scattering Green's function reads
\begin{equation} 
\begin{aligned}
G_{0}(q'',\varepsilon)=[\varepsilon-H(q'')]^{-1}.
\end{aligned}
\end{equation}

\section{Summary and overlook}

The disorder, including the local charged impurity, vacancy, crack, and the electron-electron interaction,
may leads the instability as well as the nonzero DOS in Dirac-point.
It's also found that, the modulation of the anisotropic hopping and the induced inhomogeneous charges
will leads to the semi-Dirac system which with one parabolic component and one linear component in momentum space,
and according to its special band structure, we can reasonally predict that the semi-Dirac system has a relaxation feature as well as the 
dynamical polarization between the linear Dirac system and the parabolic system which will be discussed in other place.
The chiral factor ${\bf F}_{ss'}$ (Eq.(6))
is related to the ${\rm cos}2\theta_{ss'}$
due to the bilayer character (with strong interlayer coupling),
but for weakly coupled bilayer system, the factor ${\bf F}_{ss'}$ (Eq.(6))
is still related to the ${\rm cos}\theta_{ss'}$ but not ${\rm cos}2\theta_{ss'}$.
This is also agree with the result of Ref.\cite{Gamayun O V}.
Then we can expect a new form of the chiral factor for the semi-Dirac system
which dependent on the interlayer coupling strength.
Specially, for parabolic system with Mexican hat band structure\cite{Das P},
the DOS at Fermi level $D_{F}^{bi}$ is no more energy-independent
but related to the "height" of the Mexican hat band.

We deduce the relaxation rate at low-temperature of the Dirac and Weyl system at a SiO$_{2}$ substrate,
where the elastic backscattering is dominant due to the phase-space restriction.
While at finite temperature, the anisotropic is rised due to the hole-phonon scattering\cite{Fischetti M V}
with nonparabolic effect.
The anisotropic effect (to the relaxation) also induced by the external electric field along the transport direction,
which can be treated as an asymmetric component (related to the scattering angle) of the distribution function.
However, the isotropic polarization (or the screening of the impurity scattering by the conduction electrons)
is preserved\cite{Liu Y} beyong the Thomas-Fermi approximation ${\bf q}\le 2{\bf k}_{F}$ (i.e., the scattering is constricted within
the Fermi surface).
The static polarization in long-wavelength limit is also obtained in relaxation time approximation(Eq.(24)),
which is proportional to the DOS at Fermi level,
and that's also consistent with the previous results\cite{Adam S}.
Our results are also valid for the bilayer graphene or other bilayer systems with strong interlayer coupling.

\section{Appendix.A}
The energy of the bilayer silicene $E^{bi}=\varepsilon$ can be obtained by solving Eq.(2) as
\begin{equation} 
\begin{aligned}
\varepsilon^{\pm}=\mu+M_{s}\sigma_{z}\pm\frac{1}{2}\sqrt{F_{1}},
\end{aligned}
\end{equation}
where 
\begin{equation} 
\begin{aligned}
F_{1}=&4 M_{s}^{2} + 4 \mu^{2} + 4 \lambda_{SOC}^{2} + 4 \eta \mu \lambda_{R1} \sigma_{y} + 8 M_{s} \mu \sigma_{z} + 
 4 \eta M_{s} \lambda_{R1} \sigma_{y} \sigma_{z} + 2 \eta \hbar k^{2} \tau'_{x} t_{w} v + 4 h^{2} k^{2} v_{F}^{2} - \\
& 4 E_{\perp} \eta \lambda_{SOC} \sigma_{z} \overline{\Delta} + E_{\perp}^{2} \overline{\Delta}^{2} - 
 8 a k \lambda_{R2} \lambda_{SOC} \sigma_{y} \sigma_{z} {\rm cos}\theta + 4 k \mu \tau'_{x} t_{w} {\rm cos}\theta + \\
& 2 \eta k \lambda_{R1} \sigma_{y} \tau'_{x} t_{w} {\rm cos}\theta + 4 k M_{s} \sigma_{z} \tau'_{x} t_{w} {\rm cos}\theta - 
 2 k \lambda_{R1} \sigma_{x} \tau'_{y} t_{w} {\rm cos}\theta + 8 \eta \hbar k \mu v {\rm cos}\theta + \\
& 4 \hbar k \lambda_{R1} \sigma_{y} v_{F} {\rm cos}\theta + 8 \eta \hbar k M_{s} \sigma_{z} v {\rm cos}\theta + 
 4 a E_{\perp} \eta k \lambda_{R2} \sigma_{y} \overline{\Delta} {\rm cos}\theta - \\
& 4 a^{2} k^{2} \lambda_{R2}^{2} {\rm cos}2\theta + 2 \eta \hbar k^{2} \tau'_{x} t_{w} v {\rm cos}2\theta + 
 8 a k \lambda_{R2} \lambda_{SOC} \sigma_{x} \sigma_{z} {\rm sin}\theta - 4 h k \lambda_{R1} \sigma_{x} v_{F} {\rm sin}\theta - \\
& 4 a E_{\perp} \eta k \lambda_{R2} \sigma_{x} \overline{\Delta} {\rm sin}\theta - 
 4 a^{2} k^{2} \lambda_{R2}^{2} \sigma_{x} \sigma_{y} {\rm sin}2\theta + \\
& 2 \hbar k^{2} \tau'_{y} t_{w} v_{F} {\rm sin}2\theta + (8 \eta \lambda_{SOC} \sigma_{y} \sigma_{z} \tau'_{x} + 
 8 \lambda_{SOC} \sigma_{x} \sigma_{z} \tau'_{y} - \\
&    4 E_{\perp} \sigma_{y} \tau'_{x} \overline{\Delta} - 4 E_{\perp} \eta \sigma_{x} \tau'_{y} \overline{\Delta} + 
    8 a k \lambda_{R2} (\eta \tau'_{x} - \sigma_{x} \sigma_{y} \tau'_{y}) {\rm cos}\theta + \\
&    8 a k \lambda_{R2} (\eta \sigma_{x} \sigma_{y} \tau'_{x} + \tau'_{y}) {\rm sin}\theta) \lambda_{SOC\perp} +
 8 (-1 + \eta \sigma_{x} \sigma_{y} \tau'_{x} \tau'_{y}) (\lambda_{SOC\perp})^{2} + \\
& 2 (2 \mu \tau'_{x} + \eta \lambda_{R1} \sigma_{y} \tau'_{x} + 2 M_{s} \sigma_{z} \tau'_{x} + \lambda_{R1} \sigma_{x} \tau'_{y} + 
    2 k (t_{w} + \eta \hbar \tau'_{x} v) {\rm cos}\theta - \\
&    2 \hbar k \tau'_{y} v_{F} {\rm sin}\theta) t_{\perp}.
\end{aligned}
\end{equation}
The above expression is a two-band model since we only consider the pseudospin degree of freedom.
For a simlification,
the four band model considering the spin and valley degrees of freedom,
the energy can be expressed as
\begin{equation} 
\begin{aligned}
\varepsilon^{\pm}=\pm\sqrt{m_{D}^{2}+(\frac{t^{2}_{\perp}}{2})+\hbar^{2}v_{F}^{2}k^{2}+(-1)^{a}
\sqrt{(\frac{t^{4}_{\perp}}{4})+\hbar^{2}v_{F}^{2}k^{2}\cdot(t^{2}_{\perp}+(2m_{D})^{2})}}.
\end{aligned}
\end{equation}
where $\alpha=\pm 1$.
That is similar to the bilayer graphene\cite{Tabert C J2,Prada E,AS}.
Note that, in fact the bilayer silicene has four\cite{Liu F} or more\cite{Padilha J E} stacking ways,
we only discuss the general case here.
For a detailly DFT study about the seversal kinds of bilayer silicene, see, e.g., Refs.\cite{Wu C H1,Wu C H2,Wu C H3}.

\section{Appendix.B}
The one-loop polarization function for bilayer silicene reads\cite{Gorbar E V}
\begin{equation} 
\begin{aligned}
\Pi^{bi}({\bf q},i\Omega)=-g_{s}g_{v}\frac{2\pi e^{2}T}{\epsilon_{0}\epsilon}\sum_{i\omega}
\int\frac{d^{2}k}{(2\pi)^{2}}
{\rm Tr}\left[{\bf 1}G({\bf k}+{\bf q},i\Omega+i\omega){\bf 1}G({\bf q},i\omega)\right],
\end{aligned}
\end{equation}
where the Matsubara frequencies reads $i\Omega=\Omega+i0$ with $\Omega=2m\pi T$ and $\omega=2(n+1)\pi T$,
and ${\bf 1}$ is the unit matrix.
The Green's function 
\begin{equation} 
\begin{aligned}
G({\bf q},i\omega)=
\frac{i}{{\bf 1}(\omega+i0-\mu)+{\bf k}+m_{D}}.
\end{aligned}
\end{equation}
Then in the case of strong interlayer coupling ($t_{\perp}=2\ {\rm eV}>t$),
the polarization reads\cite{Gamayun O V}
\begin{equation} 
\begin{aligned}
\Pi^{bi}({\bf q},i\Omega)=&-g_{s}g_{v}\frac{2\pi e^{2}T}{\epsilon_{0}\epsilon}
\left[
\int^{E_{{\bf k}}>\mu}\frac{d^{2}k}{2\pi^{2}}\frac{1-{\rm cos}2\theta_{ss'}}{2}
\frac{E_{{\bf k}'}+E_{{\bf k}}}{\Omega^{2}-(E_{{\bf k}'}+E_{{\bf k}})^{2}}\right.\\
&\left.+\int^{E_{{\bf k}}<\mu}\frac{d^{2}k}{2\pi^{2}}\frac{1+{\rm cos}2\theta_{ss'}}{2}
\frac{E_{{\bf k}'}-E_{{\bf k}}}{\Omega^{2}-(E_{{\bf k}}-E_{{\bf k}'})^{2}}
\right],
\end{aligned}
\end{equation}
where ${\bf k}'={\bf k}+{\bf q}$ and the term ${\rm cos}2\theta_{ss'}$ has been presented in the Eq.(7).
In strong-interlayer-coupling limit $t_{\perp}\rightarrow \infty$ with large scattering wave vector,
the parabolic spectrum has
$E_{{\bf k}}=\frac{\hbar^{2}q^{2}3a^{2}t^{2}}{4\hbar^{2}t_{\perp}}\rightarrow q^{2}/t_{\perp}$.

\end{large}
\renewcommand\refname{References}

\clearpage

\section{Tables}

Table.1: Phonon frequency in the symmetry points of monolayer silicene and MoS$_{2}$.
\begin{table}[!hbp]
\centering
\resizebox{\textwidth}{!}{
\begin{threeparttable}
\begin{spacing}{1.19}
\begin{tabular}{|c|c|c|c|}
\hline
Phonon model (silicene)               &  $\Gamma$-point                  & K-point   &   M-point            \\
\hline    
LA         &  0 cm$^{-1}$                         &  188 cm$^{-1}$   &     106 cm$^{-1}$    \\

TA        &  0 cm$^{-1}$                        & 103 cm$^{-1}$       &    100 cm$^{-1}$  \\   

ZA        &  0 cm$^{-1}$     &  103 cm$^{-1}$          &    100 cm$^{-1}$  \\
\hline       
\hline
Phonon model (MoS$_{2}$)              &  $\Gamma$-point                  & K-point   &   M-point            \\
\hline    
LA         &  0 cm$^{-1}$                         &  234 cm$^{-1}$   &     235 cm$^{-1}$    \\

TA        &  0 cm$^{-1}$                        & 186 cm$^{-1}$       &    154.6 cm$^{-1}$  \\   

ZA        &  0 cm$^{-1}$     &  176 cm$^{-1}$          &    170.8 cm$^{-1}$  \\
\hline  
\end{tabular}
\end{spacing}
\end{threeparttable}}
\end{table}
\clearpage
Fig.1
\begin{figure}[!ht]
   \centering
 \centering
   \begin{center}
     \includegraphics*[width=1\linewidth]{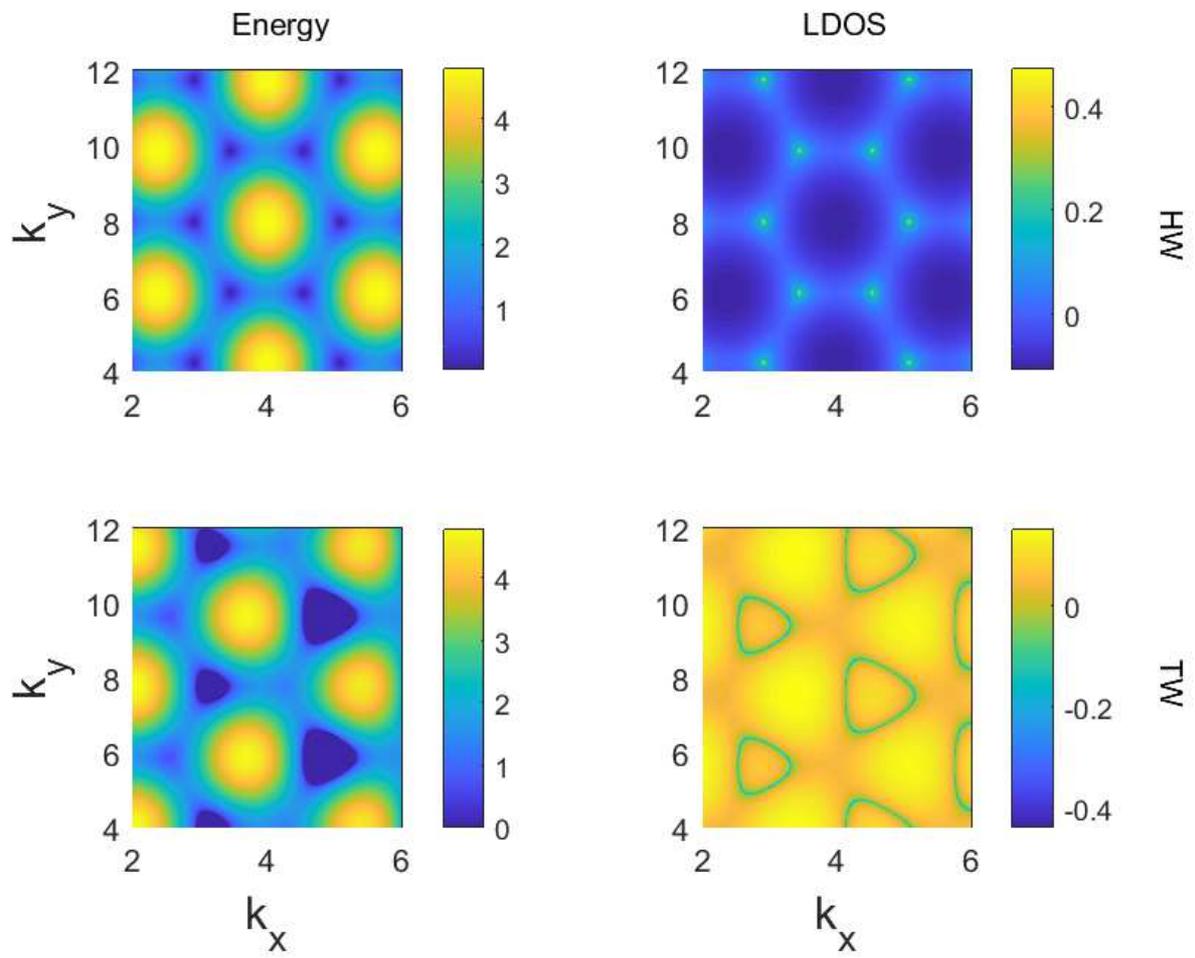}
\caption{(Color online) Energy and density plot of the monolayer silicene with hexagonal warping (upper)
and the bilayer silicene with trigonal warping (lower) in momentum space.
}
   \end{center}
\end{figure}
\clearpage
Fig.2
\begin{figure}[!ht]
   \centering
 \centering
   \begin{center}
     \includegraphics*[width=0.6\linewidth]{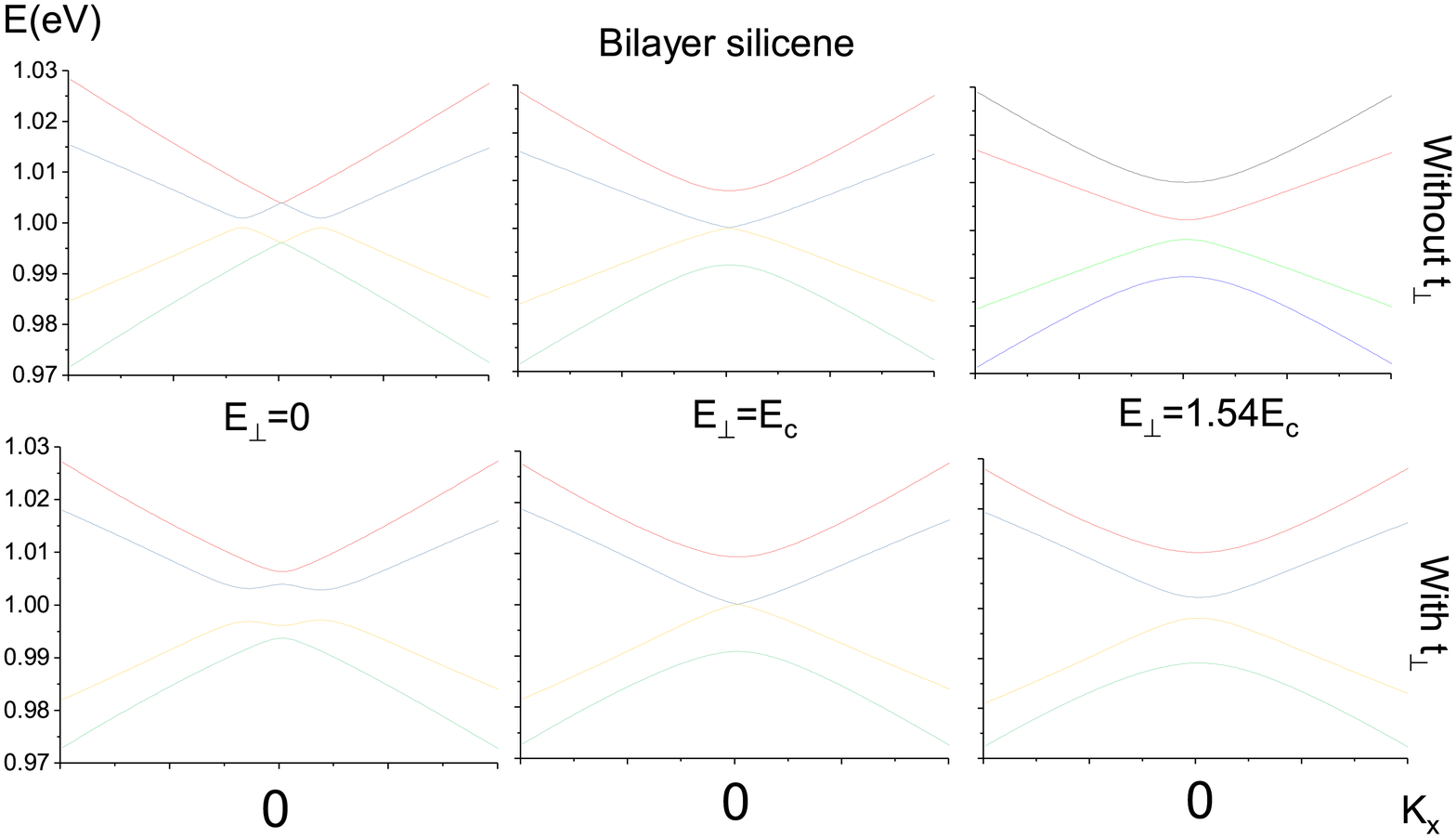}
\caption{(Color online) Band structure of he bilayer silicene under different electric field.
The upper panel without contains the effect of interlayer hopping,
while the lower panel contains the effect of interlayer hopping.
}
   \end{center}
\end{figure}
\clearpage

Fig.3
\begin{figure}[!ht]
   \centering
 \centering
   \begin{center}
     \includegraphics*[width=0.6\linewidth]{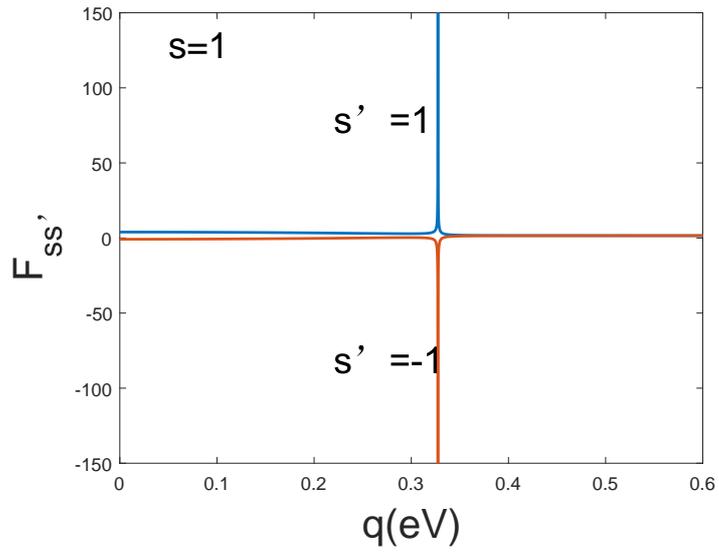}
\caption{(Color online) The azimuthal integral of the chirality factor ${\bf F}_{ss'}$ over all possible scattering angle.
}
   \end{center}
\end{figure}
\clearpage
Fig.4
\begin{figure}[!ht]
   \centering
 \centering
   \begin{center}
     \includegraphics*[width=1\linewidth]{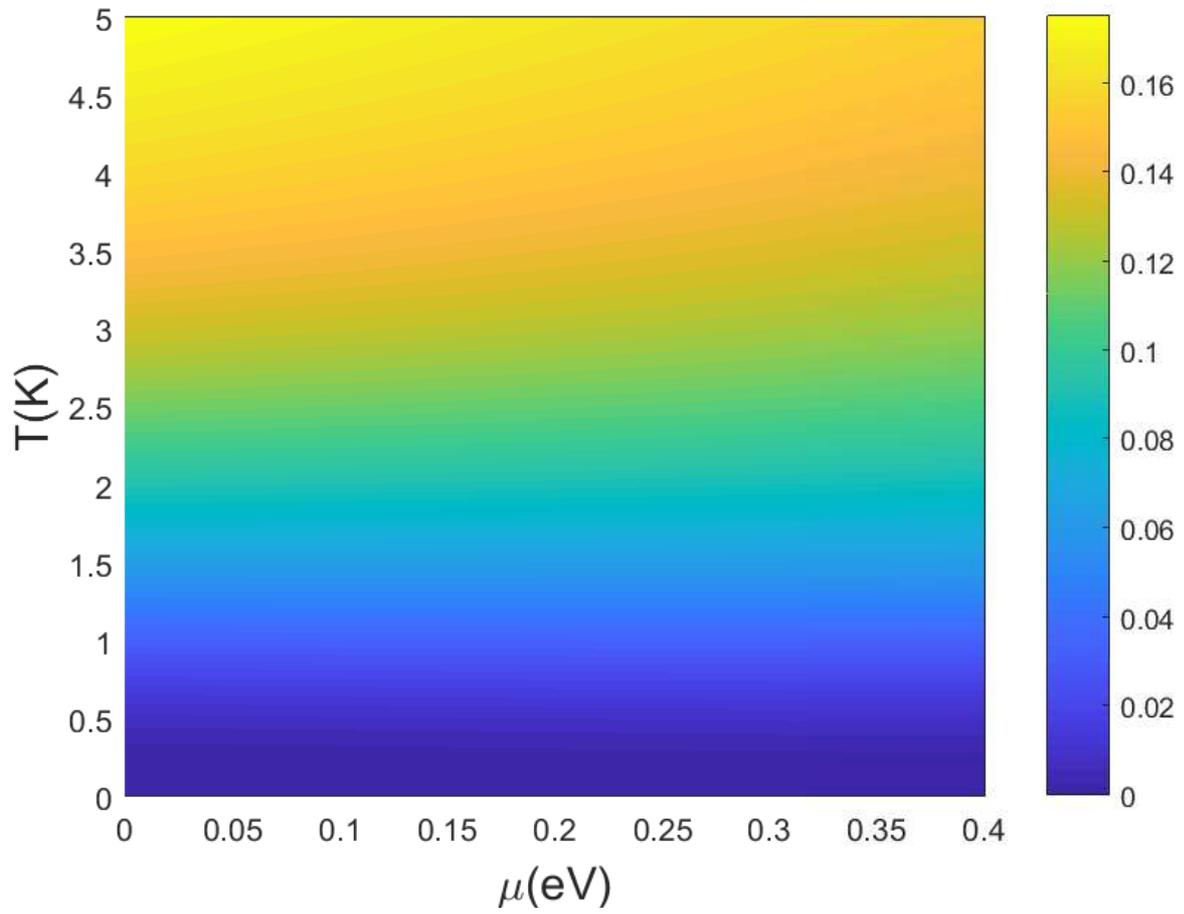}
\caption{(Color online) The residual density as a function of temperature $T$ and chemical potential $\mu$.
}
   \end{center}
\end{figure}
\clearpage
Fig.5
\begin{figure}[!ht]
   \centering
 \centering
   \begin{center}
     \includegraphics*[width=1\linewidth]{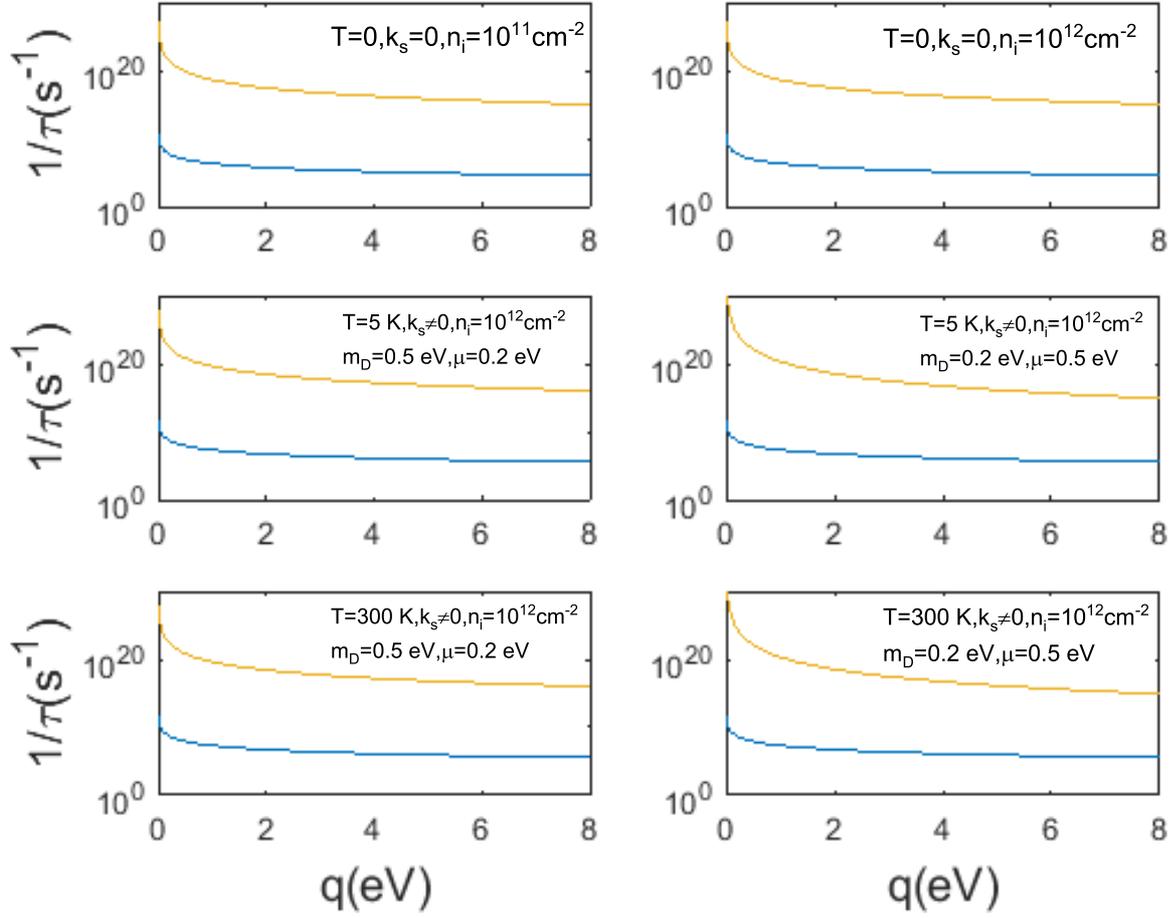}
\caption{(Color online) Relaxation time of the impurity scattering in the absence or presence of the screening wave vector
${\bf k}_{s}$.
The blue line and yellow line correspond to the 2D Dirac system and the 3D Weyl system
(in the absence of the chiral anomaly), respectively.
The Fermi velocity is setted as $5.5\times 10^{5}$ here,
the background dielectric constant is $\epsilon=2.45$,
and the corresponding impurity concentration is indicated in each panel.
}
   \end{center}
\end{figure}

\clearpage
Fig.6
\begin{figure}[!ht]
\subfigure{
\begin{minipage}[t]{0.5\textwidth}
\centering
\includegraphics[width=1\linewidth]{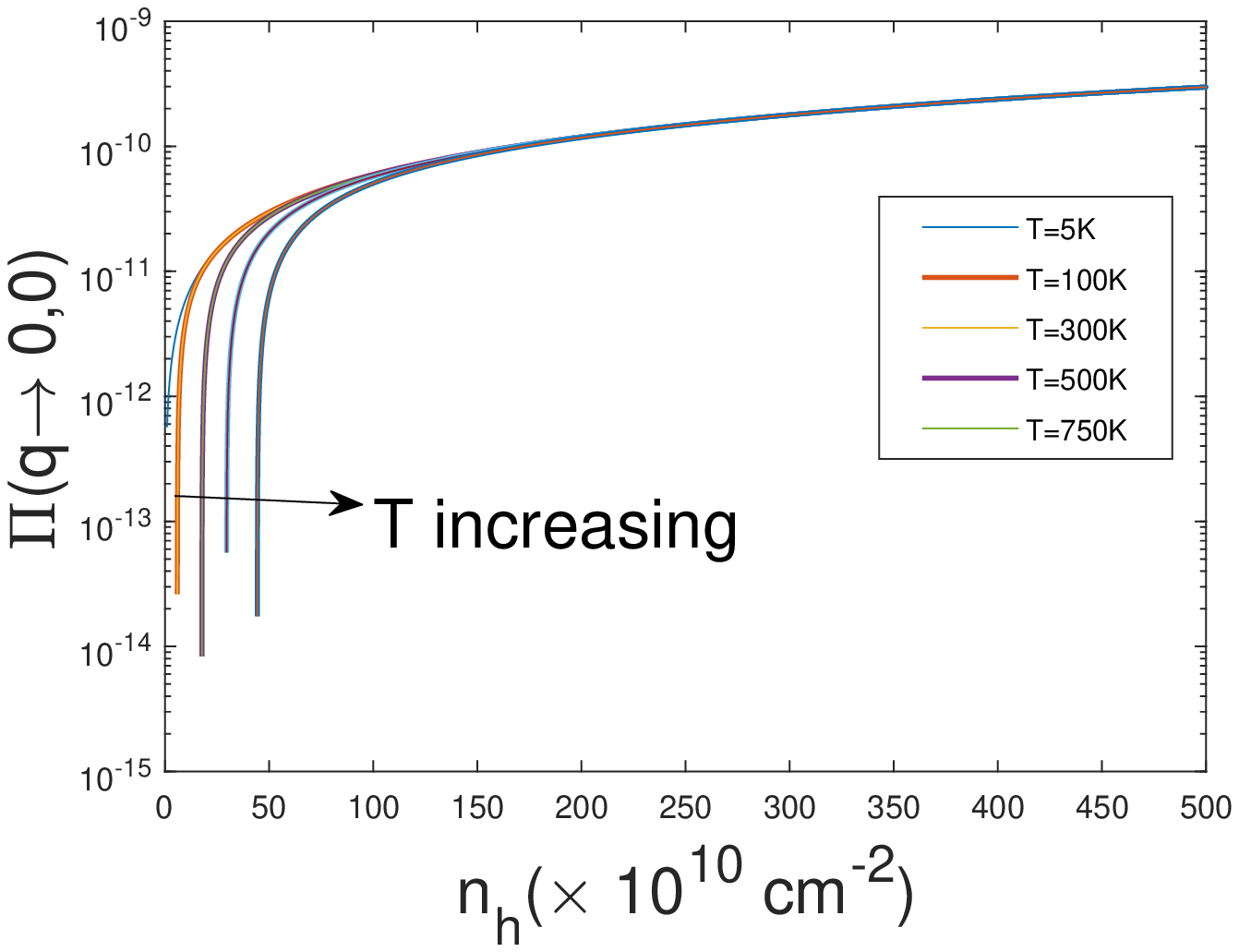}
\label{fig:side:a}
\end{minipage}
}\\
\subfigure{
\begin{minipage}[t]{0.55\textwidth}
\centering
\includegraphics[width=0.9\linewidth]{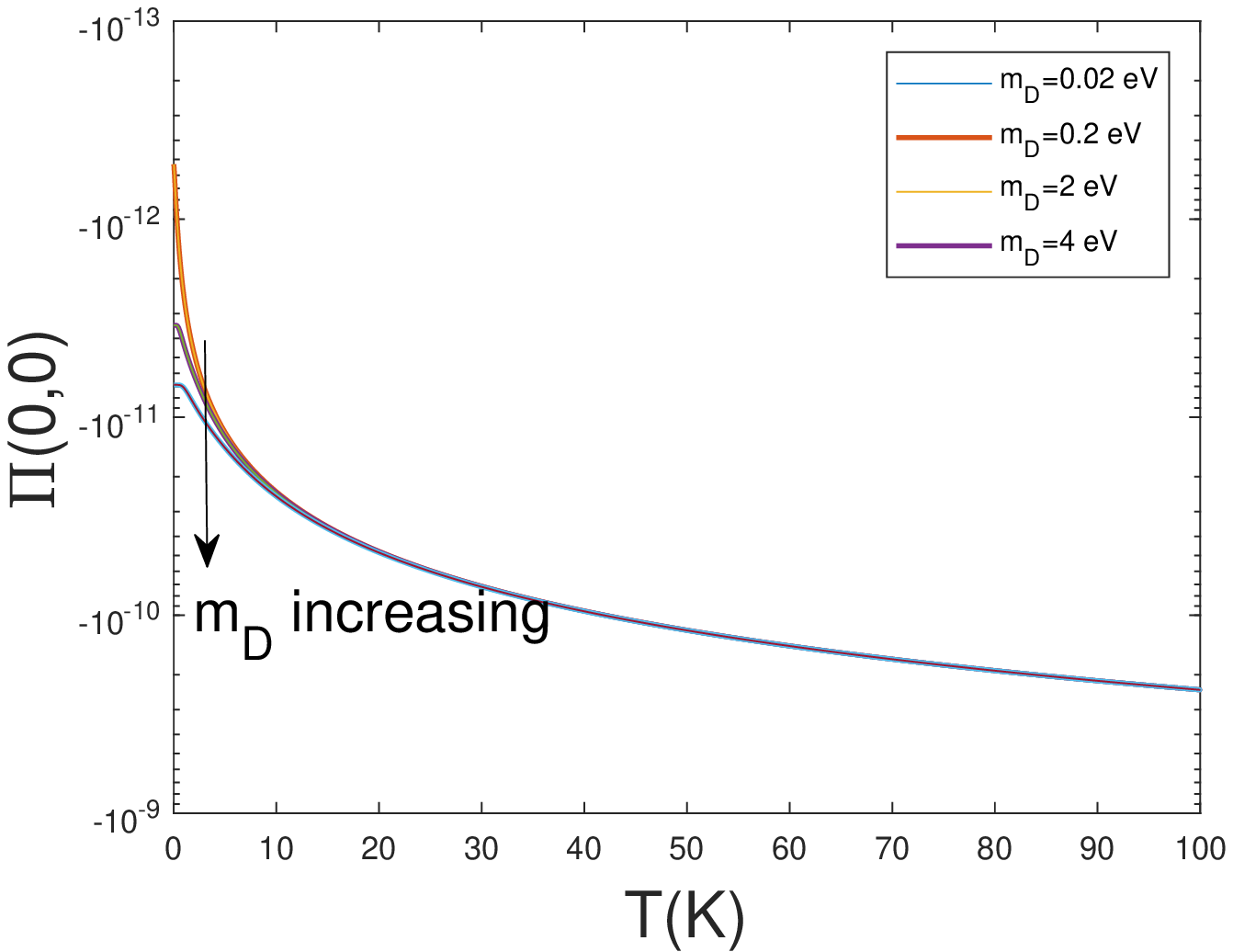}
\label{fig:side:b}
\end{minipage}
}
\caption{(Color online) Static polarization in ${\bf q}\rightarrow 0$ limit and relaxation-time approximation 
as a function of hole density $n_{h}$
with different temperature (upper),
and the $\Pi(0,0)$ as a function of temperature with different Dirac-mass
(here the chemical potential is setted as 0.2 eV).
Note that in the transverse axis of upper panel, we use the unit of $n_{h}=10^{10}$ for 
visualize the trend.
}
\end{figure}

\end{document}